\documentclass[12pt,aps]{revtex4}

\usepackage{bbm}
\usepackage{amssymb}
\usepackage{amsmath}
\usepackage{url}

\newcommand{\nc}{\newcommand}
\nc{\be}{\begin{equation}}
\nc{\ee}{\end{equation}}
\nc{\bea}{\begin{eqnarray}}
\nc{\eea}{\end{eqnarray}}
\nc{\nn}{\nonumber}
     
\nc{\lp}{\left(}
\nc{\rp}{\right)}
\catcode`\*=11  % use * as ordinary char temporarily
\def\slashsym#1#2{\mathpalette{\sl*sh{#1}}{#2}}
\def\sl*sh#1#2#3{\ooalign{\setbox0=\hbox{$#2\not$}
                          $\hfil#2\mkern-24mu\mkern#1mu
                           \raise.15\ht0\box0\hfil$\cr
                          $#2#3$}}
\catcode`\*=12

\def\pslash{{\slashsym8p}}

\def\Aslash{{\slashsym9A}}

\def\Bslash{{\slashsym6B}}

\def\Ac{{\mathcal{A}}}

\def\Fc{{\mathcal{F}}}
\def\Hc{{\mathcal{H}}}

\def\Dc{{\mathcal{D}}}

\def\Oc{{\mathcal{O}}}

\def\Det{{\mathrm{Det\,}}}

\nc{\markx}{$ \clubsuit $}
\nc{\eq}{{Eq.}}
\nc{\eqs}{{Eqs.}}
\nc{\tr}{\textrm{tr}}

\begin{document}
\rightline{HD-THEP-08-17}
\rightline{UAB-FT-656}

%\vskip 0.2in

\title{
Sizable CP Violation in the Bosonized Standard Model
}

\author{Andres Hernandez$^{(1)}$}
\email[]{A.Hernandez@thphys.uni-heidelberg.de}

\author{Thomas Konstandin$^{(2)}$}
\email[]{Konstand@ifae.es}

\author{Michael G. Schmidt$^{(1)}$}
\email[]{M.G.Schmidt@thphys.uni-heidelberg.de}

\affiliation{$(1)\,$
Institut f\"ur Theoretische Physik, Heidelberg
University, Philosophenweg 16, D-69120 Heidelberg, Germany
}%

\affiliation{$(2)\,$
Institut de F\'isica d'Altes Energies,
Edifici Cn.,
Universitat Aut\`onoma de Barcelona,
E-08193 Bellaterra (Barcelona), Spain
}

\date{\today}

\begin{abstract}
Using the worldline method, we derive an effective action of the
bosonic sector of the Standard Model by integrating out the fermionic
degrees of freedom. The CP violation stemming from the complex phase
in the CKM matrix gives rise to CP-violating operators in the one-loop
effective action in the next-to-leading order of a gradient expansion.
We calculate the prefactor of the appropriate operators and give
general estimates of CP violation in the bosonic sector of the
Standard Model. In particular, we show that the effective CP violation
for weak gauge fields is not suppressed by the Yukawa couplings of the
light quarks and is much larger than the bound given by the Jarlskog
determinant.
\end{abstract}

\maketitle

%
%%%%%%%%%%%%%%%%%%%%%%%%%%%%%%%%%%%%%%%%%%%%%%%%%%%%%%%%%%%%%%%%%%%%%%%%%%%%%%%
%  MAIN TEXT
%%%%%%%%%%%%%%%%%%%%%%%%%%%%%%%%%%%%%%%%%%%%%%%%%%%%%%%%%%%%%%%%%%%%%%%%%%%%%%%
%

%%%%%%%%%%%%%%%%%%%%%%%%%%%%%%%%%%%%%%%%%%%%%%%%%%%%%%%%%%%
\section{Introduction}
%%%%%%%%%%%%%%%%%%%%%%%%%%%%%%%%%%%%%%%%%%%%%%%%%%%%%%%%%%%%

It is a well established fact that CP violation in the Standard Model
is very small.  Main reason for this is that the sole CP-violating
effects stem from the Yukawa couplings of the quarks. In particular,
the Yukawa sector is constrained by the special flavor structure of the
Standard Model~\cite{CKMmatrix:1973,Jarlskog} suppressing CP violation. To be
explicit, the CP violation arises due to the following terms in the
Lagrangian
\be
 Y^u_{ij}  \, \bar Q_L^i \, u_R^j \phi + 
Y^d_{ij}  \, \bar Q_L^i \, d_R^j \tilde\phi +{h.c.}, 
\ee
where $Q_L$ denotes the left-handed quark $SU(2)_L$ doublet, $d_R$ and
$u_R$ denote the right-handed quark singlets and $\phi$ denotes the
Higgs doublet. We also defined the field $\tilde \phi$ by
\be
\tilde \phi = \epsilon \phi^* = 
\begin{pmatrix}
0 & -1 \\
1 & 0 \\
\end{pmatrix}
\begin{pmatrix}
\phi^0 \\
\phi^+ \\
\end{pmatrix}^*=
\begin{pmatrix}
-\phi^- \\
\phi^{0*} \\
\end{pmatrix},
\ee
and $Y^u$ and $Y^d$ denote the Yukawa coupling matrices. Under CP
conjugation, the Yukawa couplings transform as
\be
{\cal CP} \, Y^{u/d} \, {\cal CP}^{-1} = (Y^{u/d})^{*},
\ee
such that imaginary entries in $Y^{u/d}$ potentially constitute CP
violation. Spontaneous breakdown of the $SU(2)_L$ symmetry gives then
rise to the SM quark masses via
\bea
&& \hskip -1 cm Y^u_{ij}  \, \bar u_L^i \, u_R^j \, \left< \phi^0 \right> + 
Y^d_{ij}  \, \bar d_L^i \, d_R^j \, \left< \phi^0 \right> +{h.c.} \nn\\
&=& \bar u_L\, m_u \, u_R + \bar d_L\, m_d \, d_R +{h.c.}
\eea
However, not all entries in the Yukawa matrices are observable.  The
Yukawa couplings are the only terms in the SM Lagrangian that are
sensitive to global $SU(3)_R$ flavor transformations. This leads to the
conclusion that physical observables can only depend on the
combinations $m_u m_u^\dagger$ and $m_d m_d^\dagger$. In addition,
there are six global phases in the left-handed quark sector that are
unobservable in the SM.

In Ref.~\cite{Jarlskog} it was shown that in perturbation
theory the first CP odd combination of the Yukawa couplings that is
invariant under these transformations is the so-called Jarlskog
determinant
\be
\label{eq_def_Jinv}
\delta_{CP} = \textrm{Im  Det} \left[ \frac{m_u m_u^\dagger}{v^2}, 
\frac{m_d m_d^\dagger}{v^2}\right] 
= J \prod_{i<j} \frac{\tilde m_{u,i}^2 - \tilde m_{u,j}^2}{v^2}
\prod_{i<j} \frac{\tilde m_{d,i}^2 - \tilde m_{d,j}^2}{v^2}
\simeq 10^{-19},
\ee
where $\tilde m^2_{u/d}$ denote the diagonalized mass matrices
according to
\be
m_d m_d^\dagger = D \tilde m^2_d D^\dagger, \quad
m_u m_u^\dagger = U \tilde m^2_u U^\dagger.
\ee
The identity in \eq~(\ref{eq_def_Jinv}) results then from the relation
\be
\label{eq_ckm_sum}
\textrm{Im} \left[ C_{ab} C^\dagger_{bc} C_{cd} C^\dagger_{da} 
\right]
= J \sum_{e,f} \epsilon_{ace} \epsilon_{bdf}, \quad
C= U^\dagger D
\ee
(summation over indices is only performed as explicitly shown) with
the Jarlskog invariant $J$ given in terms of the standard
parametrization of the CKM matrix $C$ as \cite{Jarlskog,Yao:2006px}
\be\label{eq:valJ}
J=s_1^2s_2s_3c_1c_2c_3 \sin(\delta)=(3.0\pm0.3)\times10^{-5}.
\ee
The Jarlskog determinant in \eq~(\ref{eq_def_Jinv}) reflects the fact
that CP violation is absent if any two up-type masses or any two
down-type masses are equal. This is required since in this case there
is an additional global flavor symmetry that can be used to remove all
complex phases from the Yukawa matrices (in the SM case of three quark
families).

However, the above argument is based on the assumption that the
observable under consideration is perturbative in the Yukawa
couplings. For example, CP violation is much larger in the neutral
Kaon system than indicated by the Jarlskog determinant. If CP
violation in the mixing properties and decay rates of neutral Kaons
are considered, the CP-violating effects are suppressed by the
Jarlskog invariant $J$, but not by the Jarlskog determinant
$\delta_{CP}$. Experimentally one finds the value~\cite{Yao:2006px}
\be
\frac{\left< \pi^0 \pi^0 | {\cal H} | K_L \right>}
{\left< \pi^0 \pi^0 | {\cal H} | K_S \right>} \approx
\frac{\left< \pi^+ \pi^- | {\cal H} | K_L \right>}
{\left< \pi^+ \pi^- | {\cal H} | K_S \right>} \approx
2.2 \times 10^{-3},
\ee
which is many orders of magnitude larger than the Jarlskog
determinant. This is due to the fact that the initial and final states
in the calculation of the decay rates have a well defined quark
content and Kaons are distinct from other mesons. If e.g. the strange
and bottom quarks would be degenerate in mass, the Kaon would be
indistinguishable from the B-mesons and the CP violation in meson
decays would be non-observable. However, the quark masses are not
degenerate, and the CP violation in the Kaon system is not suppressed
by differences in Yukawa couplings as they appear in
\eq~(\ref{eq_def_Jinv}), but rather depends on ratios of Yukawa
couplings and not on the small Yukawa couplings themselves. In this
sense, CP violation in the Kaon system is a non-perturbative effect in
the quark masses and hence does not need to be suppressed by the
Jarlskog determinant~\cite{Farrar:1993hn, Ellis:1976}.

In cosmology, the main interest in CP violation originates from
baryogenesis. Sakharov pointed out~\cite{Sakharov:1967dj} that CP
violation is a prerequisite for any dynamical generation of the
observed baryon asymmetry. A baryogenesis mechanism that is based on
the SM would be most compelling~\cite{Shaposhnikov:1987,Rubakov:1996}, but this requires that the Jarlskog
determinant as an upper bound on CP violation be evaded. Even though
non-perturbative effects are obviously present in the QCD sector of
the SM, it is not expected that CP violation from the CKM matrix would
play any role in the early Universe, since a viable baryogenesis
mechanism can only operate at temperatures higher than the electroweak
scale when the sphaleron process provides the needed baryon number
violation. On the other hand, at temperatures of the electroweak
scale, the quark masses are (besides the top mass) much smaller than
the relevant energy scale and hence can be treated perturbatively. It
has been argued that in this case, the CP violation might be only
suppressed by the temperature rather than by the Higgs vev as given in
\eq~(\ref{eq_def_Jinv}), but nevertheless this would be insufficient
to be significant in a baryogenesis mechanism unless coherent scattering at a first order phase transition bubble wall and a very distinctive behaviour of the various quarks is assumed~\cite{Farrar:1993sp, Farrar:1993hn}. This created a controversial discussion~\cite{Gavela:1994}. 

In principle, there are several possibilities to avoid this dilemma
and to obtain a significant source of CP violation in the SM as
required by baryogenesis. The first option is to consider other
rephasing invariants besides the one in \eq~(\ref{eq_def_Jinv}). For
example, during a first-order phase transition, the Higgs vev changes
and hence makes it possible to construct rephasing invariants that do
not only contain the masses but also their derivatives that are
non-vanishing during the phase
transition~\cite{Konstandin:2003dx}. However, in the SM these two
quantities are proportional to each other, such that no significant
enhancement can be obtained. The second possibility is to consider
finite temperature effects that in general lead to a break-down of
perturbation theory in the infrared. This way, CP violation might be
enhanced by several orders of magnitude as demonstrated in
Ref.~\cite{Konstandin:2003dx}, but baryogenesis based on this effect
is still implausible.

Finally, CP violation can be considered in the context of effective
actions. Consider the SM at low energies with gauge fields that are
weak compared to the energy scale of the quark masses. If the
fermionic degrees of freedom are integrated out, a purely bosonic
theory describes the physics at low energies. In this case, the CP
violation in the quark sector will eventually give rise to higher
dimensional operators as first proposed in Ref.~\cite{Smit}. In the present
work we will demonstrate that, different from the leading order case~\cite{Smit}, 
in the next-to-leading order of the
gradient expansion, the effective action indeed contains CP violation
that exceeds the perturbative bound given in
\eq~(\ref{eq_def_Jinv}). Main motivation for this approach is the
scenario of cold electroweak baryogenesis
\cite{Shaposhnikov:1999,Smit:2003,Smit} that specifically utilizes
lattice simulations of the bosonic sector of the SM with higher
dimensional operators that violate the CP symmetry.

The approach is based on the determination of the covariant current
using the worldline method as presented in Ref.~\cite{worldline}. An
alternative and more direct method was recently proposed in
Ref.~\cite{Salcedo:2008bs}. Even though this direct method nicely
avoids the matching of the action to the current, the method we use
still exhibits some computational advantages. In particular, the
worldline method does not involve momentum integrations, avoids the
handling of the $\gamma$ matrix algebra, and is easily implemented
with a computer algebra program.

The paper is organized as follows: In Sec.~\ref{sec_smit} we review
the effective action in leading order of the gradient expansion. In
Sec.~\ref{sec_next} the next-to-leading order effective action is
discussed before we conclude in Sec.~\ref{sec_concl}. In Appendix
\ref{app} we comment on some aspects of the explicit form of the
effective action at next-to-leading order.

%%%%%%%%%%%%%%%%%%%%%%%%%%%%%%%%%%%%%%%%%%%%%%%%%%%%%%%%%%%
\section{The Effective Action at Leading Order\label{sec_smit}}
%%%%%%%%%%%%%%%%%%%%%%%%%%%%%%%%%%%%%%%%%%%%%%%%%%%%%%%%%%%

In this section, we present the leading order of the effective action
as first presented in Ref.~\cite{Salcedo1} and also derived in
Ref.~\cite{worldline} using the worldline method
\cite{Schmidt:1993,Schmidt:1994,Schmidt:1995}. Besides, we discuss the
absence of CP violation at this order following Ref.~\cite{Smit}.

Consider the Euclidean Dirac operator
\begin{equation}
\label{O}
\Oc \equiv \pslash - i \Phi(x) - \gamma_5 \Pi(x) - \Aslash(x)
- \gamma_5 \Bslash(x),
\end{equation}
where the external fields have a general internal group structure,
e.g.~a flavor or gauge matrix structure. We are interested in the imaginary part of the one-loop
effective action that contains the CP-violating contributions to the
action
\be
W^{-}= \arg\left(\Det[\Oc]\right).
\ee
As shown elegantly in Ref.~\cite{Gagne}, the imaginary part of the
action can be reformulated in terms of variables that have a
well-defined behavior under chiral transformations, namely
\be
\Ac_{\mu}=
\begin{pmatrix}
A_{\mu}^L & 0 \cr 0 & A_{\mu}^R \\
\end{pmatrix}=
\begin{pmatrix}
A_{\mu}+B_{\mu} & 0 \cr 0 & A_{\mu}-B_{\mu} \\
\end{pmatrix}, \;
\ee
\be
\Hc =
\begin{pmatrix}
0 & i\, H \\
-i\, H^{\dagger} & 0 \\
\end{pmatrix}=
\begin{pmatrix}
0 & i\, \Phi + \Pi \\
-i\, \Phi + \Pi & 0 \\
\end{pmatrix}.
\ee
The fermions under consideration are the quarks of the SM, such that
the gauge fields belong to the $SU(3)_c \times SU(2)_L \times U(1)_Y$
gauge group. Since the color interactions are not essential for CP
violation, we suppress any $SU(3)$ indices. Besides a gauge index, the
fields also carry a flavor index. In particular, the
scalar/pseudo-scalar background field is of the form
\be
H=
\begin{pmatrix}
\phi^0 & \phi^+ \\
\phi^- & -\phi^{0*} \\
\end{pmatrix}
\begin{pmatrix}
Y_u & 0 \\
0 & -Y_d \\
\end{pmatrix},
\ee
where $Y_{u/d}$ denote the SM Yukawa coupling matrices.

Since the effective action is gauge invariant, we still have the
freedom to simplify the action by a certain choice of gauge. As
detailed in Ref.~\cite{Smit} a convenient choice is the unitary gauge,
in which the Higgs field is of the form
\be
H= \phi^0(x_\mu)
\begin{pmatrix}
Y_u & 0 \\
0 & Y_d \\
\end{pmatrix}=
\begin{pmatrix}
m_u & 0 \\
0 & m_d \\
\end{pmatrix}.
\ee
In addition, we perform a basis transformation that diagonalizes the
mass terms $m_u$ and $m_d$. This transformation is not compatible with
$SU(2)_L$ gauge invariance, such that the gauge invariance in the
resulting expression is realized non-linearly. In this basis, the
$SU(2)_L$ gauge field strength is then of the following form in
flavor space
\be
\Fc_L = 
\begin{pmatrix} 
F_0 & F^+ C \\
C^\dagger F^- & -F_0 \\
\end{pmatrix},
\ee
where $C$ denotes the CKM matrix as defined in \eq~(\ref{eq_ckm_sum}).

The effective action is most compactly presented in the labeled
operator notation that was introduced in Ref.~\cite{Salcedo1}, and
used also in Ref.~\cite{worldline}. In this notation, mass matrices
obtain an additional subscript that indicates the position of the mass
matrix in a subsequent product of operators. For example, using
this notation we write
\be
m_1 m^3_2 m^2_3 \, \Dc_{\mu}\Hc \, \Dc_{\nu}\Hc =
m \, \Dc_{\mu}\Hc \, m^3 \,  \Dc_{\nu}\Hc \, m^2.
\ee
A detailed definition and applications of this notation can be found
in Ref.~\cite{Salcedo1} and we refer the reader to this work. Using
this notation, the chiral invariant part of the leading order
contribution in the gradient expansion has to be of the form
\bea
\label{eq_action_lo}
W_{lo}^{-}&=& \epsilon^{\mu\nu\lambda\sigma} 
\bigl\langle i N(m_1, m_2, m_3) \, \Dc_{\mu}\Hc \, \Dc_{\nu}\Hc \, 
\Fc_{\lambda\sigma} \nn \\
&& + N(m_1, m_2, m_3, m_4) \, 
\Dc_{\mu}\Hc \, \Dc_{\nu}\Hc \, \Dc_{\lambda}\Hc \,
\Dc_{\sigma}\Hc \bigr\rangle,
\eea
with some functions $N(m_1, m_2, m_3)$ and $N(m_1, m_2, m_3, m_4)$.

In order to contribute to CP violation, an expression has to contain
at least four CKM matrices. In this case, the arguments that lead to
the relation in \eq~(\ref{eq_ckm_sum}) can be used to extract the
CP-violating parts.  Applying these considerations to the expression
in \eq~(\ref{eq_action_lo}) implies that
\begin{itemize}

\item 
The term proportional to $(\Dc\Hc)^2 \, \Fc$ does not contribute since
it contains at most three CKM matrices.

\item 
The term proportional to $(\Dc\Hc)^4$ contains four CKM matrices. In
this case all four operators have to be left-handed and charged,
i.e. it is proportional to
\bea
&& \hskip -1 cm (\Dc\Hc^2)_L^+ (\Dc\Hc^2)_L^- (\Dc\Hc^2)_L^+ (\Dc\Hc^2)_L^- \nn\\
&& \propto (m_2^2 - m_1^2)(m_3^2 - m_2^2)(m_4^2 - m_3^2)(m_1^2 + m_4^2)
 A^+_L A^-_L A^+_L A^-_L  , 
\eea
where the subscripts and superscripts $L,+,-$ denote which parts have been
projected out in terms of transformation properties under chiral and
$U(1)_\textrm{em}$ transformations.

\item 
The contraction of the Lorentz indices of this term with the
Levi-Civita $\epsilon$ tensor vanishes.

\end{itemize}

We conclude, in agreement with \cite{Smit,worldline}, that the leading
order of the effective action in the gradient expansion does not
contain any CP violation in the SM. However, in next-to-leading order,
the action will contain terms like $(\Dc\Hc)^2 \,
\Fc^2$ that do not necessarily vanish after contraction with the Levi-Civita
tensor.

%%%%%%%%%%%%%%%%%%%%%%%%%%%%%%%%%%%%%%%%%%%%%%%%%%%%%%%%%%%
\section{The Effective Action at Next-To-Leading Order\label{sec_next}}
%%%%%%%%%%%%%%%%%%%%%%%%%%%%%%%%%%%%%%%%%%%%%%%%%%%%%%%%%%%

In this section we discuss some general properties of the effective
action in next-to-leading order. First, notice that if
written in terms of the gauge field $A_\mu$ and the field strength
$F_{\mu\nu}$, the coefficient of the effective action has negative
mass dimension. Hence, in the limit of vanishing masses, the effective
action becomes infinite. This is not surprising, since the gradient
expansion assumes
\be
\label{eq_applicability}
A_\mu \ll m, \quad F_{\mu\nu} \ll m^2.
\ee
This leads to the question what is the range of applicability of our
result. In order to discuss this question, we analyze the CP-violating
part of a specific term in the effective action. Consider a term of
the form
\be
R(m_1, m_2, m_3, m_4) 
\Dc_{\alpha}\Hc \, \Dc_{\alpha}\Hc \, \Fc_{\mu\nu} \, \Fc_{\lambda\sigma}.%+{h.c.}
\ee
This could in principle contain CP violation if all appearing gauge
fields are left-handed and charged after symmetry breaking. This
yields the contributions
\bea
&& \hskip -1 cm
\bar R (m^d_1, m^u_2, m^d_3, m^u_4) \, 
A_\alpha^+ \, A_\alpha^- \, F^+_{\mu\nu} \, F^-_{\lambda\sigma} \nn\\
&& + \, \bar R (m^u_1, m^d_2, m^u_3, m^d_4) \, 
A_\alpha^- \, A_\alpha^+ \, F^-_{\mu\nu} \, F^+_{\lambda\sigma},
%+ {h.c.}, 
\eea
where we used the symmetrization
\be
\bar R(m_1, m_2, m_3, m_4) = \frac{1}{16}\sum_{n_i \in \pm m_i} 
R( n_1, n_2, n_3, n_4)
(n_2 - n_1) (n_3 - n_2).
\ee
The symmetrization ensures that all appearing gauge fields are
left-handed. Changing to the mass eigenbasis and using
\eq~(\ref{eq_ckm_sum}) this can be recast as
\be
 C_1 \, A_\alpha^+ \, A_\alpha^- \, F^+_{\mu\nu} \, F^-_{\lambda\sigma}
+ \, C_2 \, A_\alpha^- \, A_\alpha^+ \, F^-_{\mu\nu} \, F^+_{\lambda\sigma}, 
%+ {h.c.},
\ee
where we use the definitions
\bea
C_1 &=& J \, \sum_{i,k,m \in \textrm{up}} \, \sum_{j,l,n \in
\textrm{down}} \, \epsilon_{ikm} \epsilon_{jln} 
\bar R (\tilde m^d_k, \tilde m^u_l, \tilde m^d_m, \tilde m^u_n ), \\
C_2 &=& - J \, \sum_{i,k,m \in \textrm{up}} \, \sum_{j,l,n \in
\textrm{down}} \, \epsilon_{ikm} \epsilon_{jln} 
\bar R (\tilde m^u_l, \tilde m^d_k, \tilde m^u_n, \tilde m^d_m ).
\eea
The subscript indicates hereby the quark flavor, $\textrm{up} = \{
u,c,t\}$ and $\textrm{down} = \{ d,s,b\}$.

Notice that this expression vanishes if two up-type masses or two
down-type masses coincide as required. However, the coefficient can be
much larger than the Jarlskog determinant stated in
\eq~(\ref{eq_def_Jinv}) even in units of the light quark masses
$\tilde m^{-2}_{u/d}$. The largest contribution results typically from
the contribution involving only the four lightest quarks.

%[\markx I adopted the following part to our results] 
Let us come back
to the question of the range of applicability of the gradient
expansion. In principle, one would expect that the largest
contributions be proportional to $\tilde m^{-2}_{u/d}$ or even
larger, e.g.  $\tilde m^2_{c/b} \tilde m^{-4}_{u/d}$. In this case,
the mass scale that indicates the breakdown of the gradient expansion
in \eq~(\ref{eq_applicability}) would be given by the lightest quarks
invalidating the gradient expansion already for very weak external
fields. Besides, there might be one more obstacle, namely the physical
infrared divergences of the light quarks. The operator under
consideration describes a scattering process that is indistinguishable
from the same process including a soft quark/anti-quark pair. Hence,
the amplitudes can contain contributions that scale as $\log{\tilde
m^2_{u}}$ or $\log{\tilde m^2_{d}}$ in the massless limit. This 
would require that the corresponding operators 
with soft quarks in the initial/final states be taken into account .

Fortunately, it turns out that all appearing CP-violating
contributions are finite in the limit of vanishing up/down quark
masses and there are only terms that scale as $\Oc\left(\tilde
m^{-2}_{c},\tilde m^{-2}_{b},\tilde m^{-2}_{t}\right)$. We hence
expect that the range of validity in
\eq~(\ref{eq_applicability}) is at least given by the scale of the
charm quark mass.

In fact, the range of applicability can be even larger according to
the following argument. For simplification, imagine that there is a
common energy scale for the gradient expansion
\be
A_\mu \sim  \partial_\mu \sim E, \quad 
F_{\mu\nu} \sim \partial^2_\mu  \sim E^2.
\ee
In the limit of weak fields $E \ll \tilde m_c$ we obtain the
estimate for CP violation in the effective action
\be
\label{eq_est_low}
W^- \propto J \, \tilde m_c^{-2} E^6,
\ee
while in the case of a strong background, $E \gg \tilde m_t$, the
effective action could be expanded in the quark masses. In this case,
following the argument by Jarlskog, one obtains on dimensional grounds
an estimate for CP violation similar to the Jarlskog determinant,
namely
\be
\label{eq_est_high}
W^- \propto J \, \tilde m^4_t \tilde m_b^4 \tilde m^2_c \tilde m_s^2 E^{-8}.
\ee
Comparison of these two limits indicates that the transition region is
given for energies
\be
E \sim (m^4_t \tilde m_b^4 \tilde m^4_c \tilde m_s^2)^{1/14} \simeq 5.0
\textrm{ GeV}.
\ee
and below this value the effective action presented here should
indicate the correct order of magnitude of CP violation in the bosonic
sector of the SM.

Using the method developed in Ref.~\cite{worldline} we calculated the
effective action explicitly. The specific form of the coefficient
functions is too large to be presented here, but they are available as
computer files
\footnote{The complete imaginary part of the effective action can be
found at \texttt{http://www.thphys.uni-heidelberg.de/\textasciitilde
schmidt/Weff\_nlo/}}. Appendix \ref{app} contains some more general
comments on the action and its coefficient functions.

Interestingly, almost all the contributions cancel amongst themselves,
and there is only one contribution to the CP-violating part of the
effective action, namely
\bea\label{eq:CPterm}
\frac{1}{8(4\pi)^2}\frac{3}{16} \frac{J\,\kappa^{CP}}{\tilde m_c^2}
\epsilon^{\mu\nu\lambda\sigma}\int d^4x
\biggl(Z_{\mu}W^{+}_{\nu\lambda}W^{-}_{\alpha}
\left(W^{+}_{\sigma}W^{-}_{\alpha}+W^{+}_{\alpha}W^{-}_{\sigma}\right)
+ \, c.c.\biggr)
\eea
with $J$ given by Eq.~(\ref{eq:valJ}) and
\be\label{kappa}
\kappa^{CP} \approx 9.87.
\ee

Finally, notice that the action can always be rewritten in $SU(2)_L$
gauge invariant quantities. For example, the charged gauge fields can
be rewritten as
\be
W_{\mu\nu}^+ = \frac{\phi^\dagger W_{\mu\nu} \tilde \phi}{ \phi^\dagger \phi}, \quad
W_{\mu\nu}^- = \frac{{\tilde \phi}^\dagger W_{\mu\nu} \phi}{ \phi^\dagger \phi}, \quad
W_{\mu}^+ = \frac{\phi^\dagger {\cal D}_\mu \tilde \phi}{ \phi^\dagger \phi}, \quad
W_{\mu}^- = \frac{{\tilde \phi}^\dagger {\cal D}_{\mu} \phi}{ \phi^\dagger \phi}, \quad
\ee
and similarly for the uncharged quantities
\be
Z_\mu = W_\mu^3 - B_\mu = \frac{\phi^\dagger {\cal D}_{\mu} \phi  
- {\tilde \phi}^\dagger {\cal D}_{\mu} \tilde\phi }
{ 2\phi^\dagger \phi}, \quad
h^{-1} \partial_\mu h = \frac{\phi^\dagger {\cal D}_{\mu} \phi  
+ {\tilde \phi}^\dagger {\cal D}_{\mu} \tilde\phi }
{ 2\phi^\dagger \phi},
\ee
and 
\be
W_{\mu\nu}^3 = \frac{\phi^\dagger W_{\mu\nu} \phi}
{ \phi^\dagger \phi}.
\ee
%%%%%%%%%%%%%%%%%%%%%%%%%%%%%%%%%%%%%%%%%%%%%%%%%%%%%%%%%%%
\section{Conclusions\label{sec_concl}}
%%%%%%%%%%%%%%%%%%%%%%%%%%%%%%%%%%%%%%%%%%%%%%%%%%%%%%%%%%%

We calculated the CP-violating contributions to the effective action
in the bosonized Standard Model in next-to-leading order in the
gradient expansion. Surprisingly after some cancelations only one term remained, given in Eq.~(\ref{eq:CPterm}), our main result. 
We argued that the resulting action should be
valid for bosonic fields whose energy scale does not exceed much the charm
mass. This observation is based on the fact that the action after IR
regularization remains finite in the limit of vanishing up and down
quark masses. We find that the coefficients of the resulting
dimension-six operators are suppressed by the charm mass and the
Jarlskog invariant $J$ but are many orders larger than the Jarlskog
determinant $\delta_{CP}$. It will be interesting to see the results for cold
electroweak baryogenesis following the lines of \cite{Smit:2003}. In principle 
the temperature enters as an additional mass scale into the
calculation. It should be possible to derive an expansion of the effective
action that is valid for external fields whose energy scale
exceeds the charm mass but not the temperature and that is non-perturbative in the quark masses. 
This issue is under further study.

%Accordingly, the effective action can e.g. be used to study the
%influence of rather large instantons (with radius $R \gtrsim 1/\tilde
%m_c$) on the bosonized Standard Model. The impact of sphalerons cannot
%reliably be deduced from the effective action presented because
%sphalerons act at high temperature, in which case the

%%%%%%%%%%%%%%%%%%%%%%%%%%%%%%%%%%%%%%%%%%%%%%%%%%%%%%%%%%%
\section*{Acknowledgments}
%%%%%%%%%%%%%%%%%%%%%%%%%%%%%%%%%%%%%%%%%%%%%%%%%%%%%%%%%%%

T.K. is supported by the EU FP6 Marie Curie Research \& Training
Network 'UniverseNet' (MRTN-CT-2006-035863).  A.H. is supported by
CONACYT/DAAD, Contract No.~A/05/12566.

%%%%%%%%%%%%%%%%%%%%%%%%%%%%%%%%%%%%%%%%%%%%%%%%%%%%%%%%%%%
\appendix
%%%%%%%%%%%%%%%%%%%%%%%%%%%%%%%%%%%%%%%%%%%%%%%%%%%%%%%%%%%

%%%%%%%%%%%%%%%%%%%%%%%%%%%%%%%%%%%%%%%%%%%%%%%%%%%%%%%%%%%
\section{Effective Action and CP-violating Contributions \label{app}}
%%%%%%%%%%%%%%%%%%%%%%%%%%%%%%%%%%%%%%%%%%%%%%%%%%%%%%%%%%%
The imaginary part of the effective action in four dimensions in
next-to-leading order in a gradient expansion takes the form
\bea\label{eq:ansatzW}
W_{nlo}^{-}&=&\epsilon^{\mu\nu\lambda\sigma}\biggr\langle
+\frac{1}{4}Q^{(1)}_{123}\Dc_{\alpha}\Fc_{\mu\nu}\Fc_{\lambda\sigma}\Dc_{\alpha}\Hc
+\frac{1}{4}Q^{(2)}_{123}\Dc_{\alpha}\Fc_{\mu\nu}\Fc_{\lambda\alpha}\Dc_{\sigma}\Hc\nonumber\\
&&\hspace{1.2cm}
+\frac{1}{4}Q^{(4)}_{123}\Dc_{\alpha}\Fc_{\mu\alpha}\Fc_{\nu\lambda}\Dc_{\sigma}\Hc
+\frac{i}{2}Q^{(5)}_{123}\Dc_{\alpha}\Dc_{\alpha}\Dc_{\mu}\Hc\Fc_{\nu\lambda}\Dc_{\sigma}\Hc\nonumber\\
&&\hspace{1.2cm}
+\frac{i}{2}R^{(6)}_{1234}\Fc_{\mu\nu}\Dc_{\alpha}\Dc_{\alpha}\Hc\Dc_{\lambda}\Hc\Dc_{\sigma}\Hc
+\frac{i}{2}R^{(7)}_{1234}\Fc_{\mu\nu}\Dc_{\lambda}\Hc\Dc_{\alpha}\Dc_{\alpha}\Hc\Dc_{\sigma}\Hc\nonumber\\
&&\hspace{1.2cm}
+\frac{1}{4}R^{(9)}_{1234}\Fc_{\mu\nu}\Fc_{\lambda\alpha}\Dc_{\sigma}\Hc\Dc_{\alpha}\Hc
+\frac{1}{4}R^{(10)}_{1234}\Fc_{\mu\nu}\Dc_{\lambda}\Hc\Fc_{\sigma\alpha}\Dc_{\alpha}\Hc\nonumber\\
&&\hspace{1.2cm}
+\frac{i}{2}R^{(12)}_{1234}\Fc_{\mu\nu}\Dc_{\alpha}\Dc_{\lambda}\Hc\Dc_{\sigma}\Hc\Dc_{\alpha}\Hc
+\frac{i}{2}R^{(13)}_{1234}\Fc_{\mu\nu}\Dc_{\lambda}\Hc\Dc_{\alpha}\Dc_{\sigma}\Hc\Dc_{\alpha}\Hc\nonumber\\
&&\hspace{1.2cm}
+\frac{1}{4}R^{(14)}_{1234}\Fc_{\mu\nu}\Dc_{\alpha}\Hc\Fc_{\lambda\sigma}\Dc_{\alpha}\Hc
+\frac{1}{4}R^{(15)}_{1234}\Fc_{\mu\alpha}\Fc_{\nu\alpha}\Dc_{\lambda}\Hc\Dc_{\sigma}\Hc\nonumber\\
&&\hspace{1.2cm}
+\frac{i}{2}S^{(1)}_{12345}\Fc_{\mu\nu}\Dc_{\lambda}\Hc\Dc_{\sigma}\Hc\Dc_{\alpha}\Hc\Dc_{\alpha}\Hc
+\frac{i}{2}S^{(2)}_{12345}\Fc_{\mu\nu}\Dc_{\lambda}\Hc\Dc_{\alpha}\Hc\Dc_{\sigma}\Hc\Dc_{\alpha}\Hc\nonumber\\
&&\hspace{1.2cm}
+\frac{i}{2}S^{(3)}_{12345}\Fc_{\mu\nu}\Dc_{\lambda}\Hc\Dc_{\alpha}\Hc\Dc_{\alpha}\Hc\Dc_{\sigma}\Hc
+\frac{i}{2}S^{(4)}_{12345}\Fc_{\mu\nu}\Dc_{\alpha}\Hc\Dc_{\lambda}\Hc\Dc_{\sigma}\Hc\Dc_{\alpha}\Hc\nonumber\\
&&\hspace{1.2cm}
+S^{(7)}_{12345}\Dc_{\alpha}\Dc_{\mu}\Hc\Dc_{\nu}\Hc\Dc_{\lambda}\Hc\Dc_{\sigma}\Hc\Dc_{\alpha}\Hc
+S^{(8)}_{12345}\Dc_{\alpha}\Dc_{\mu}\Hc\Dc_{\nu}\Hc\Dc_{\lambda}\Hc\Dc_{\alpha}\Hc\Dc_{\sigma}\Hc\nonumber\\
&&\hspace{1.2cm}
+T^{(1)}_{123456}\Dc_{\mu}\Hc\Dc_{\nu}\Hc\Dc_{\lambda}\Hc\Dc_{\sigma}\Hc\Dc_{\alpha}\Hc\Dc_{\alpha}\Hc
+T^{(2)}_{123456}\Dc_{\mu}\Hc\Dc_{\nu}\Hc\Dc_{\lambda}\Hc\Dc_{\alpha}\Hc\Dc_{\sigma}\Hc\Dc_{\alpha}\Hc\nonumber\\
&&\hspace{1.2cm}
+T^{(3)}_{123456}\Dc_{\mu}\Hc\Dc_{\nu}\Hc\Dc_{\alpha}\Hc\Dc_{\lambda}\Hc\Dc_{\sigma}\Hc\Dc_{\alpha}\Hc
\biggr\rangle+h.c.
\eea 
Due to the trace, partial integration and other possible
manipulations, the expression for $W_{nlo}^{-}$ in
\eq~(\ref{eq:ansatzW}) is not unique. Through a judicious set of such
transformations, $W_{nlo}^{-}$ was brought into a simpler form than
the one obtained originally from the matching procedure, into one
which is finite term by term at all coincidence limits. The superscripts of the
functions distinguish hereby between different coefficient functions
with the same number of arguments. The superscripts are not
consecutively numbered what is reminiscent of the fact that we
obtained this action by removing some contributions of a more general
ansatz.

The explicit functions are not shown here for space considerations,
but in order to give the reader an impression of their form, we
present the simplest function that is given by
\begin{align}
Q^{(2)}_{123}=&\frac{8}{\left(3 (m_1^2-m_2^2)^2 (m_1+m_2) (m_2^2-m_3^2)^2 (m_1+m_3) (m_2+m_3)\right)}\times\nn\\
&\bigl( 
m_1^4 \left(m_2^2-m_2 m_3+m_3^2\right) \left(m_2^2+4 m_2 m_3+m_3^2\right)
+m_2^4 m_3 \left(2 m_2^3-5 m_2^2 m_3+m_3^3\right)
\nn\\
&
+m_1^3 m_2 m_3 (m_2+m_3) \left(3 m_2^2-2 m_2 m_3+3 m_3^2\right)\nn\\
&
+m_1 m_2^3 (m_2+m_3) \left(2 m_2^3-9 m_2^2 m_3+3 m_3^3\right)\nn\\
&
+m_1^2 m_2^2 \left(-5 m_2^4-9 m_2^3 m_3+11 m_2^2 m_3^2+m_2 m_3^3-2 m_3^4\right)
\bigr)\nn\\
&
+\frac{8 m_1^3 \left(m_1^4+m_1^3 m_2-3 m_1^2 m_2^2+3 m_1 m_2^3+6 m_2^3 m_3\right) \log\left[\frac{m_1^2}{m_2^2}\right]}
{3 (m_1^2-m_2^2)^3 (m_1+m_2) (m_1^2-m_3^2) (m_1+m_3)}\nn\\
&
-\frac{8 m_3^3 \left(6 m_1 m_2^3+m_3 \left(3 m_2^3-3 m_2^2 m_3+m_2 m_3^2+m_3^3\right)\right) \log\left[\frac{m_2^2}{m_3^2}\right]}{3 (m_1^2-m_3^2) (m_1+m_3) (m_2^2-m_3^2)^3 (m_2+m_3)}.
\end{align}
All the other functions, while increasing in complexity as the number
of arguments increases, are of this form: rational functions of the
masses, eventually multiplied by logarithms of mass ratios. In
particular, all functions are homogeneous in their arguments for
dimensional reasons
\begin{equation}
Q(a\,m_1,a\,m_2,a\,m_3)=\frac{1}{a^2}Q(m_1,m_2,m_3).
\end{equation}
The $Q$ functions, lacking sufficient CKM matrices, cannot contribute
CP-violating terms. Therefore the CP-violating terms can only appear
from the $R$, $S$ and $T$ functions. As mentioned in the main text, almost all
the expressions cancel. Essentially just one of the contributions coming from
$R^{(12)}$, $R^{(13)}$ and their conjugates survive, see Eq.~(\ref{eq:CPterm}). In calculating
the coefficient in Eq.~(\ref{kappa}) we used the full analytic
functions, but the final result is too big to present it
here. However, in the limit where $\tilde{m}_u \to
\tilde{m}_d \to 0$ and $\tilde{m}_b \to \tilde{m}_c$ the result
is simpler, and differs by around 1\% from the one given in
Eq.~(\ref{kappa}). In this limit the contribution takes the following
form
\bea
\frac{\kappa^{CP}}{\tilde{m}_c^2} &\approx&\frac{32}{9 \tilde{m}_c^2 \left(\tilde{m}_c^2-\tilde{m}_s^2\right)^3 \left(\tilde{m}_c^2-\tilde{m}_t^2\right)^3 \left(\tilde{m}_s^2-\tilde{m}_t^2\right)^2}\times\nn\\
&& \biggl(
\tilde{m}_s^6 \tilde{m}_t^6 \left(\tilde{m}_s^2-\tilde{m}_t^2\right)^2
+3 \tilde{m}_c^{14} \left(\tilde{m}_s^2+\tilde{m}_t^2\right)\nn\\
&&
-5 \tilde{m}_c^2 \tilde{m}_s^4 \tilde{m}_t^4 \left(\tilde{m}_s^2-\tilde{m}_t^2\right)^2 \left(\tilde{m}_s^2+\tilde{m}_t^2\right)
-12 \tilde{m}_c^{12} \left(\tilde{m}_s^4+\tilde{m}_t^4\right)\nn\\
&&
+\tilde{m}_c^4 \tilde{m}_s^2 \tilde{m}_t^2 \left(\tilde{m}_s^2-\tilde{m}_t^2\right)^2 \left(13 \tilde{m}_s^4+28 \tilde{m}_s^2 \tilde{m}_t^2+13 \tilde{m}_t^4\right)
+18 \tilde{m}_c^{10} \left(\tilde{m}_s^6+\tilde{m}_t^6\right)\nn\\
&&
+\tilde{m}_c^8 \left(-12 \tilde{m}_s^8+37 \tilde{m}_s^6 \tilde{m}_t^2-74 \tilde{m}_s^4 \tilde{m}_t^4+37 \tilde{m}_s^2 \tilde{m}_t^6-12 \tilde{m}_t^8\right)\nn\\
&&
+\tilde{m}_c^6 \left(3 \tilde{m}_s^{10}-41 \tilde{m}_s^8 \tilde{m}_t^2+41 \tilde{m}_s^6 \tilde{m}_t^4+41 \tilde{m}_s^4 \tilde{m}_t^6-41 \tilde{m}_s^2 \tilde{m}_t^8+3 \tilde{m}_t^{10}\right)\biggr)\nn\\
&&
-\frac{64\, \tilde{m}_c^4 \tilde{m}_s^2 \tilde{m}_t^2 \left(\tilde{m}_c^2-\tilde{m}_t^2\right) \left(\tilde{m}_c^2-3 \tilde{m}_s^2+2 \tilde{m}_t^2\right) \log\left[\frac{\tilde{m}_s^2}{\tilde{m}_c^2}\right]}{3 \left(\tilde{m}_c^2-\tilde{m}_s^2\right)^4 \left(\tilde{m}_s^2-\tilde{m}_t^2\right)^3}\nn\\
&&
+\frac{64\, \tilde{m}_c^4 \tilde{m}_s^2 \left(\tilde{m}_c^2-\tilde{m}_s^2\right) \tilde{m}_t^2 \left(\tilde{m}_c^2+2 \tilde{m}_s^2-3 \tilde{m}_t^2\right) \log\left[\frac{\tilde{m}_t^2}{\tilde{m}_c^2}\right]}{3 \left(\tilde{m}_c^2-\tilde{m}_t^2\right)^4 \left(\tilde{m}_s^2-\tilde{m}_t^2\right)^3}.
\eea

\end{document}